\documentclass[12pt]{article}
\usepackage{graphicx}
\begin{document}

\title{Competing orders in high-$T_{c}$ superconductors}
\author{Guo-Zhu Liu$^{1,3}$ and Geng Cheng$^{2,3}$ \\
$^{1}${\small {\it Lab of Quantum Communication and Quantum Computation, }}\\
{\small {\it University of Science and Technology of China, }}\\
{\small {\it Hefei, Anhui, 230026, P.R. China }}\\
$^{2}${\small {\it CCAST (World Laboratory), P.O. Box 8730, Beijing 100080,
P.R. China }}\\
{\small {\it $^{3}$Department of Astronomy and Applied Physics, }}\\
{\small {\it University of Science and Technology of China, }}\\
{\small {\it Hefei, Anhui, 230026, P.R. China}}}
\maketitle

\begin{abstract}
\baselineskip20pt Within a (2+1)-dimensional U(1) gauge field
theory, after calculating the Dyson-Schwinger equation for fermion
self-energy we find that chiral symmetry breaking (CSB) occurs if
the gauge boson has a very small mass but is suppressed when the
mass is larger than a critical value. In the CSB phase, the
fermion acquires a dynamically generated mass, which leads to
antiferromagnetic (AF) long-range order. Since in the
superconducting (SC) state the gauge boson acquires a finite mass
via Anderson-Higgs mechanism, we obtain a field theoretical
description of the competition between the AF order and the SC
order. As a compromise of this competition, there is a coexistence
of these two orders in the bulk material of cuprate
superconductors.
\end{abstract}

\newpage
\baselineskip20pt

Understanding the competition of various ground states of cuprate
superconductors is one of the central problems in modern condensed
matter physics. At hall-filling, the cuprate superconductor is a
Mott insulator with antiferromagnetic (AF) long-range order. When
the doping concentration increases, AF order rapidly disappears
and superconducting (SC) order emerges as the ground state. How to
describe this evolution from AF order to SC order upon doping in a
simple way is a very interesting problem.

Since the discovery of high temperature superconductors, extensive
theoretical and experimental work appeared to investigate the
possibility of spin-charge separation [1,2], which states that the
spin and charge degrees of freedom might be carried by different
quasiparticles, i.e. charge carrying holons and spin carrying
spinons. The recently observed breakdown of the Wiedemann-Franz
(WF) law in underdoped cuprates [3] confirms its existence [4]. In
spin-charge separated theories, the pseudogap in the normal state
is the spin gap arising from pairing of spinons. Superconductivity
is achieved when the holons Bose condense into a macroscopic
quantum state, which indicates that the vortex of high temperature
superconductors carries a double flux quantum $hc/e$. However, so
far experiments found only $hc/2e$ vortices [5]. Recently Lee and
Wen [6] showed that SU(2) slave-boson theory [7] can naturally
lead to a stable $hc/2e$ vortex inside which a finite pseudogap
exists [8].

In this Letter, we propose that the competition between the AF
order and the SC order can be modeled by a competition between
chiral symmetry breaking (CSB) and the mass of a gauge boson
within an effective theory of the SU(2) formulation. After
calculating the Dyson-Schwinger (DS) equation for fermion
self-energy, we find that CSB happens if the gauge boson mass is
zero or very small but is destroyed when the gauge boson mass is
larger than a finite critical value. We then show that CSB
corresponds to the formation of long-range AF order by calculating
AF correlation function. Because the gauge boson mass is generated
via the Anderson-Higgs mechanism in the superconducting state,
there is a competition between AF order and SC order.

We start our discussion from the staggered flux state in the SU(2)
slave-boson treatment of the $t$-$J$ model [7]. In this paper we
adopt the following effective model of the underdoped cuprates
that consists of massless fermions, bosons and a U(1) gauge field
[4]
\begin{equation}
{\cal L}_{F}=\sum_{\sigma=1}^{N}\overline{\psi}_{\sigma}
v_{\sigma, \mu} \left( \partial_{\mu }-ia_{\mu} \right)
\gamma_{\mu}\psi_{\sigma}+\left|\left(\partial_{\mu}-ia_{\mu}
\right)b\right|^{2}+V(\left|b\right|^{2}).
\end{equation}
The Fermi field $\psi_{\sigma}$ is a $4\times 1$ spinor. The $4
\times 4$ $\gamma _{\mu}$ matrices obey the algebra, $\lbrace
\gamma_{\mu},\gamma_{\nu} \rbrace=2\delta_{\mu \nu}$, and for
simplicity, we let $v_{\sigma, \mu}=1$ ($\mu,\nu=0,1,2$).
$b=(b_{1},b_{2})$ is a doublet of scalar fields representing the
holons.

In Lagrangian (1) the fermions are massless, so it respects the
chiral symmetries $\psi \rightarrow \exp (i\theta \gamma
_{3,5})\psi $, with $\gamma _{3}$ and $\gamma _{5}$ two $4 \times
4$ matrices that anticommute with $\gamma _{\mu}$ ($\mu=0,1,2$).
If the fermion flavor is below a critical number $N_{c}$ the
strong gauge field can generate a finite mass for the fermions
[9-11], which breaks the chiral symmetries. This phenomenon is
called chiral symmetry breaking (CSB), which has been studied for
many years in particle physics as a possible mechanism to generate
fermion mass without introducing annoying Higgs particles.
Previous study found that CSB happens when the holons are absent
[9-11] and when the holons are not Bose condensed [4].

We now would like to consider the superconducting phase where
boson $b$ acquires a nonzero vacuum expectation value, i.e.,
$\langle b \rangle \neq 0$. The nonzero $\langle b \rangle$
spontaneously breaks gauge symmetry of the theory and the gauge
boson acquires a finite mass $\xi$ via Anderson-Higgs mechanism.
CSB is a low-energy phenomenon because (2+1)-dimensional U(1)
gauge field theory is asymptotically free and only in the infrared
region the gauge interaction is strong enough to cause fermion
condensation. This requires the fermions be apart from each other.
In the superconducting state, the gauge boson becomes massive and
can not mediate a long-range interaction. Intuitively, a finite
gauge boson mass is repulsive to CSB which is achieved by the
formation of fermion-anti-fermion pairs. To determine whether CSB
still occurs in the SC state, quantitative calculations should be
carried out.

CSB is a nonperturbative phenomenon and can not be obtained within
any finite order of the perturbation expansion. The standard
approach to this problem is to solve the self-consistent DS
equation for the fermion self-energy. The inverse fermion
propagator is written as $S^{-1}(p)=i\gamma \cdot p A \left( p^{2}
\right)+\Sigma \left( p^{2} \right)$, $A(p^{2})$ is the
wave-function renormalization and $\Sigma(p^{2})$ the fermion
self-energy. If the DS equation has only trivial solutions, the
fermions remain massless and the chiral symmetries are not broken.
However, not all nontrivial solutions correspond to a dynamically
generated fermion mass [12]. If a nontrivial solution of the DS
equation satisfy an additional squarely integral condition
[12,13], it then signals the appearance of CSB.

The DS equation is
\begin{equation}
\Sigma(p^{2})=\int\frac{d^{3}k}{(2\pi)^{3}}\frac{\gamma^{\mu}D_{\mu\nu}(p-k)\Sigma(k^{2})\gamma^{\nu}}{k^{2}
+\Sigma^{2}(k^{2})}.
\end{equation}
In this paper, we use the following propagator of the massive
gauge boson
\begin{equation}
D_{\mu\nu}(p-k)=\frac{8}{(N+1)(\left|p-k\right|+\eta)}\left(
\delta_{\mu\nu}-\frac{(p-k)_{\mu}(p-k)_{\nu}}{(p-k)^{2}}\right),
\end{equation}
where $\eta$ reflects the effect of the gauge boson mass $\xi$
($\eta=8\xi^{2}/(N+1)$). After performing angular integration and
introducing an ultraviolet cutoff $\Lambda$, we have
\begin{eqnarray}
\Sigma(p^{2})&=&\lambda\int^{\Lambda}_{0} dk\frac{k\Sigma(k^{2})}{k^{2}+\Sigma^{2}(k^{2})} \nonumber \\
&&\times \frac{1}{p}\left(p+k-\left|p-k\right|-\eta\ln\left(\frac{p+k+\eta}{\left|p-k\right|+\eta}\right)\right),
\end{eqnarray}
where $\lambda=4/(N+1)\pi^{2}$ serves as an effective coupling
constant. Here, for simplicity, we assumed that $A(p^{2})=1$. For
$\eta=0$, this assumption leads to $N_{c}=32/\pi^{2}$ [9]. More
careful treatment [10] calculated the DS equation for
$\Sigma(p^{2})/A(p^{2})$ considering higher-order corrections and
found that the critical behaviour is qualitatively unchanged.
Since assuming that $A(p^{2})=1$ can significantly simplify the
calculations and the higher-order corrections are small we expect
the result derived from DS equation (4) is reliable.

If we do not introduce an ultraviolet cutoff ($\Lambda \rightarrow
\infty$), the critical behavior of Eq.(4) is completely
independent of $\eta$, which can be easily seen by making the
scale transformation, $p \rightarrow p/\eta$, $k \rightarrow
k/\eta$ and $\Sigma \rightarrow \Sigma/\eta$. This scale
invariance is destroyed by an ultraviolet cutoff $\Lambda$ which
is natural because the theory (1) was originally defined on
lattices. Once an ultraviolet cutoff is intrdoduced, the solution
$\Sigma(p^{2})$ then automatically satisfies the squarely
integrable condition [13] and hence is a symmetry breaking
solution. Theoretical analysis implies that the critical fermion
number $N_{c}$ of Eq.(4) should depend on $\Lambda/\eta$.
Actually, we have showed that when the gauge boson has a very
large mass, say $\eta\gg\Lambda$, the DS equation has no physical
solutions. If the gauge boson is massless, the last term in the
kernel of Eq.(4) can be dropped off, leaving an equation that has
a critical number $32/\pi^{2}$. However, at present we do not have
a detailed dependence of the critical number $N_{c}$ on
$\Lambda/\eta$. In particular, it is not clear whether $N_{c}$ is
a monotonous function of $\Lambda/\eta$ or not. To settle this
problem, we should solve the DS equation quantitatively.

The most intriguing property of the DS equation is that it is a
(or a set of) nonlinear integral equation which exhibits many
interesting phenomena and at the same time is very hard to be
studied analytically and numerically. In this paper, based on
bifurcation theory we are able to obtain the phase transition
point of the nonlinear DS equation by studying the eigenvalue
problem for its associated Fr\^{e}chet derivative. This scheme
[14,15] simplifies the numerical computation and also can lead to
a reliable bifurcation point. We will calculate the eigenvalues of
the linearized integral equation using parameter imbedding method,
which can eliminate the uncertainty that is unavoidable when
carrying out numerical computation in the vicinity of a
singularity. The details of the computation will be given
elsewhere.

Making Fr\^{e}chet derivative of the nonlinear equation (4), we
have the following linearized equation
\begin{equation}
\Sigma(p^{2})=\lambda\int^{\Lambda/\eta}_{0}
dk\Sigma(k^{2})\frac{1}{pk}\left(p+k-\left|p-k\right|-\ln\left(\frac{p+k+1}{\left|p-k\right|+
1}\right)\right)
\end{equation}
where for calculational convenience we made the transformation $p
\rightarrow p/\eta$, $k \rightarrow k/\eta$ and $\Sigma
\rightarrow \Sigma/\eta$. The smallest eigenvalue $\lambda_{c}$ of
this equation is just the bifurcation point from which a
nontrivial solution of the DS equation (4) branches off. For
$\lambda > \lambda_{c}$, the DS equation has nontrivial solutions
and CSB happens. The ultravoilet cutoff $\Lambda$ is provided by
the lattice constant and hence is fixed. We can obtain the
relation of $N_{c}$ and $\eta$ by calculating the critical
coupling $\lambda_{c}$ for different values of $\Lambda/\eta$.

In order to get the smallest eigenvalue $\lambda_{c}$, we first
use parameter imbedding method [14,15] to convert Eq.(5) into two
differential-integral equations with $\lambda$ their variable.
After choosing an appropriate contour in the complex
$\lambda$-plane and integrating with respect to the parameter
$\lambda$, we finally obtain the exact eigenvalue $\lambda_{c}$.

Our numerical result is presented in Fig.(1). The most important
result is that the critical fermion number $N_{c}$ is a
monotonously increasing function of $\Lambda/\eta$. It conforms
our expectation that a finite mass of the gauge boson is repulsive
to CSB. For small $\Lambda/\eta$ the critical number $N_{c}$ is
smaller than physical fermion number 2, so CSB does not happen.
When $\Lambda/\eta$ increases, the critical number $N_{c}$
increases accordingly and finally becomes larger than 2 at about
$\Lambda/\eta_{c}=10^{8}$. When $\Lambda/\eta$ continues to
increase, $N_{c}$ increases more and more slowly and finally
approaches a constant value 2.15. Thus we can conclude that CSB
takes place when the gauge boson mass is zero and very
small but is destroyed when the gauge boson mass is larger than a
critical value.

We next show that CSB corresponds to the formation of AF
long-range order [16] by calculating the AF spin correlation
function. At the mean field level, the AF correlation function is
defined as [17]
\begin{equation}
\langle S^{+}S^{-} \rangle_{0}=-\frac{1}{4}\int\frac{d^{3}k}{(2\pi)^{3}}Tr\left[G_{0}(k)G_{0}(k+p)\right],
\end{equation}
where $G_{0}(k)$ is the fermion propagator. For massless fermions,
$G_{0}(k)=\frac{-i}{\gamma_{\mu} k^{\mu}}$, we have
\begin{equation}
\langle S^{+}S^{-} \rangle_{0}=-\frac{\left|p\right|}{16},
\end{equation}
and the AF correlation is largely lost. This result is natural if
we look back to our starting point, i.e., the staggered flux mean
field phase which is based on the resonating valence bond (RVB)
picture proposed by Anderson [1]. The RVB state is actually a
liquid of spin singlets, so it has only short range AF
correlation. Since a N\'{e}el order was observed in experiments,
we should find a way to get back the long-range spin correlation.
One possible way is to go beyond the mean field treatment and
include the gauge fluctuations which is necessary to impose the no
double occupation constraint. As discussed above, when strong
gauge interaction is taken into account, the system undergoes a
chiral instability and the massless fermions acquire a finite
mass. Although the dynamically generated fermion mass depends on
the 3-momentum, here, for simplicity, we assume a constant mass
$\Sigma_{0}$ for the fermions. This approximation is valid because
we only care about the low-energy property and $\Sigma(0)$ is
actually a constant. Then the propagator for the massive fermions
is written as
\begin{equation}
G_{0}(k)=\frac{-i\gamma_{\mu}k^{\mu}}{k^{2}+\Sigma_{0}^{2}},
\end{equation}
which leads to
\begin{equation}
\langle S^{+}S^{-}\rangle_{0}=-\frac{1}{4\pi}\left(\Sigma_{0}
+\frac{p^{2}+4\Sigma_{0}^{2}}{2\left|p\right|}
\arcsin\left(\frac{p^{2}}{p^{2}+4\Sigma_{0}^{2}}\right)^{1/2}\right).
\end{equation}
This spin correlation behaves like $-\Sigma_{0}/2\pi$ as $p
\rightarrow 0$ and we have long-range AF correlation when CSB
happens.

We should emphasize that CSB is necessary for producing AF
long-range order. If we only include gauge fluctuations into the
staggered spin correlation (6) while keeping the fermions
massless, then [17]
\begin{equation}
\langle
S^{+}S^{-}\rangle_{GF}=-\frac{8}{12\pi^{2}(N+1)}\left|p\right|\ln
\left(\frac{\Lambda^{2}}{p^{2}}\right)
\end{equation}
which approaches zero at the limit $p \rightarrow 0$. Wen and
coworkers [17] used to claim that long-range AF correlation can be
obtained by reexponentiating the spin correlation function based
on the conclusion that the gauge field can not generate a finite
mass for fermions and hence is a marginal perturbation. This
result is derived by means of perturbation expansion. However, CSB
is a nonperturbative phenomenon and whether the gauge field
generates a finite mass for the massless fermions can only be
settled by investigating the self-consistent DS equation for the
fermion self-energy. Studies of the DS equation show that gauge
field is strong enough to generate a mass for fermions, so it is
not a marginal perturbation. Moreover, the AF long-range order
breaks the rotational symmetry accompanying a massless Goldstone
boson (spin wave). This is difficult to understand if the AF order
is induced by a marginally perturbative gauge field, rather than
by a spontaneous symmetry breaking.

In the superconducting state, the gauge boson mass $\xi$ is
proportional to the superfluid density $\rho_{s}$, i.e.,
$\xi\sim\rho_{s}$. Since $\eta\sim\xi^{2}$ and the superfluid
density in high-$T_{c}$ superconductors is proportional to doping
concentration $\delta$, we have $\eta\sim\delta^{2}$. Then whether
CSB exists at $T\rightarrow 0$ depends on the doping level and the
knob that tunes the different orders is the holon degree of
freedom. At half-filling and very low $\delta$ CSB occurs because
the holons are not Bose condensed and the gauge boson is massless.
As $\delta$ increases the cuprate becomes a superconductor which
gives the gauge boson a finite mass. When $\delta$ is larger than
a critical value $\delta_{c}$ the mass of the gauge boson
($\xi_{c}\sim\delta_{c}$) is large enough to suppress CSB. Thus we
obtain a competition between the AF order, which dominates at
half-filling and low $\delta$, and the SC order, which dominates
at high $\delta$. As a compromise of this competition, there is a
coexistence of these two orders in the bulk material of cuprate
superconductors for doping concentration between $\delta_{c}$ and
the critical point $\delta_{sc}$ at which superconductivity begins
to emerge.

It was generally claimed that the AF order is destroyed by the
moving holes because no long-range AF correlation has been
observed in cuprates at high doping $\delta$. However, our result
[4] indicated that CSB and hence the AF order can coexist with
free holons. It seems to us that the AF order is actually
destroyed by Bose condensation of holons (or spontaneous gauge
symmetry breaking) at low temperature and thermal fluctuations at
high temperature (above $T_{c}$). To find out the true mechanism
that destroys the AF order, experiments, preferably scanning
tunnelling microscopy (STM) or neutron scattering, should be
performed at the $T \rightarrow 0$ limit when the
superconductivity is suppressed, for example, by strong magnetic
fields.

Based on spin-charge separation and CSB, we now have a clear
picture of the evolution of zero temperature ground states upon
increasing the doping concentration. For doping concentration less
than $\delta_{c}$, CSB occurs and leads to long-range AF
correlation, which is then destroyed by superconducting order for
doping concentration higher than $\delta_{c}$. On the other hand,
when superconductivity is completely suppressed by strong magnetic
fields, CSB occurs in underdoped cuprates and gives the nodal
spinons (originally gapless due to the $d$-wave symmetry of the
spin-gap) a finite mass which provides a gap for free fermions to
be excited at low temperatures, causing the breakdown of the WF
law [4]. Thus the combination of spin-charge separation and CSB
gives a unified field theoretical description of both the
breakdown of WF law and the competition between long-range AF
order and long-range SC order. This is the most notable advantage
of our scenario.

Recently, much experimental [18,19] and theoretical [20] effort
has been made to the magnetic field induced AF order in the vortex
state of cuprate superconductors. In particular, neutron
scattering and STM experiments found that AF order appears around
the vortex cores where the superfluid is suppressed by magnetic
field locally. Our result is consistent with these experiments.
Qualitatively, the AF correlation is enhanced in regions where the
superfluid density and hence the gauge boson mass becomes smaller
than their critical value. Within our scenario, since CSB can
coexist with a small gauge boson mass, the length scale for AF
order to appear should be larger than the vortex scale, which is
consistent with STM experiments [19]. However, to quantitatively
explain the experimental data, several subtle problems should be
made clear, especially the ratio of gauge boson mass to superfluid
density and the detailed distribution of the superfluid density in
the vortex state, which are beyond the scope of this paper.

We thank Cheng Lee for his help in numerical calculations and V.
P. Gusynin for pointing out one typing error in the original
manuscript. This work is supported by National Science Foundation
in China No.10175058.

\begin{figure}
\centering \includegraphics{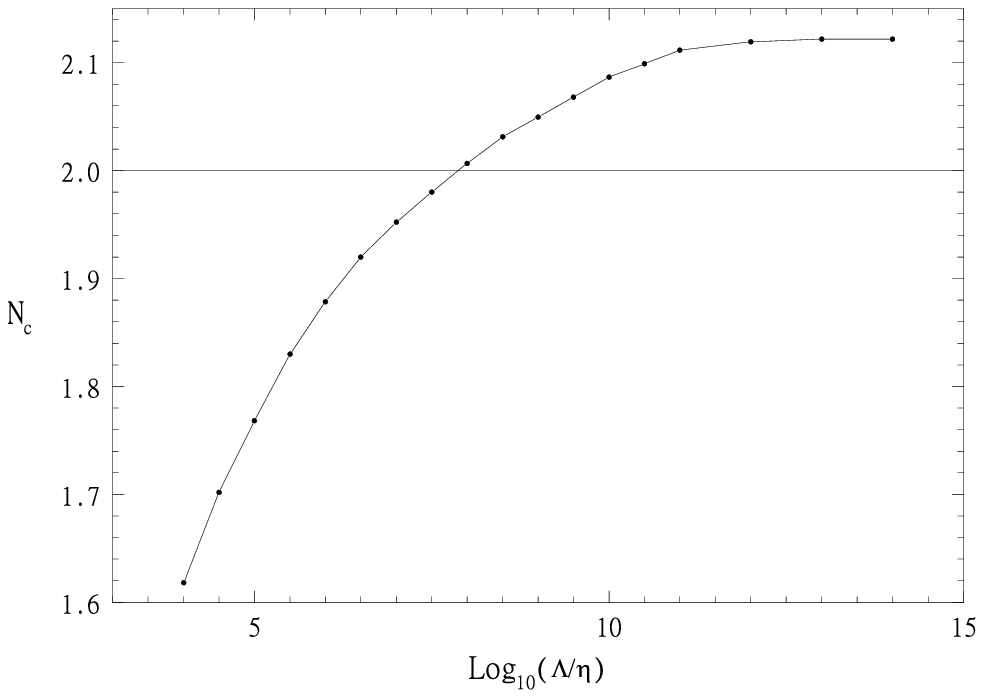}
\begin{minipage}{12cm}
\caption{The dependence of the critical number $N_{c}$ on
$\log_{10}(\Lambda/\eta)$.}
\end{minipage}

\end{figure}

\end{document}